\documentclass[amsmath,amssymb,twocolumn]{revtex4}

\usepackage[dvips]{graphicx}

\begin{document}

\title{Weak (anti-)localization in doped $Z_2$-topological insulator}

\author{Ken-Ichiro Imura, Yoshio Kuramoto, Kentaro Nomura}
\affiliation{Department of Physics, Tohoku University, Sendai 980-8578, Japan}

\begin{abstract}
Localization properties of the doped 
$Z_2$-topological insulator are studied by weak localization theory. 
The disordered Kane-Mele model for graphene is taken 
as a prototype, and analyzed with 
attention to effects of the topological mass term,
inter-valley 
scattering, and the Rashba spin-orbit interaction.
The known tendency of graphene to anti-localize in the {\it absence} of 
inter-valley scattering between $K$ and $K'$ points
is naturally placed as the massless limit of Kane-Mele model.  
The latter is shown to have a unitary behavior even in the absence of magnetic field
due to the topological mass term.
When inter-valley scattering is introduced, 
the topological mass term 
leaves the system in the unitary class, whereas 
the ordinary mass term, which appears if A and B sublattices are inequivalent, 
turns the system to weak localization.
The Rashba spin-orbit interaction 
in the {\it presence} of 
$K$-$K'$ scattering drive the system to weak anti-localization
in sharp contrast to the ideal graphene case.
\end{abstract}

\date{\today}

\maketitle

\section{Introduction}

The concept of $Z_2$ topological insulator was first introduced in a model for graphene
\cite{Geim},
in the presence of both intrinsic and extrinsic (Rashba)
spin-orbit interactions
(called hereafter, Kane-Mele model) \cite{KM1,KM2}.
The origin of $Z_2$ symmetry lies naturally in the time reversal invariance 
of the underlying spin-orbit interactions, i.e., in the existence of Kramers pairs.
In the continuum limit, the intrinsic spin-orbit interaction is represented by a
so-called topological mass term (of size $\Delta$, opening a spin-orbit gap $2\Delta$), 
encoding quantized spin Hall effect.
The latter occurs when Fermi energy is in the gap, and
implies the existence of a pair of counter-propagating gapless states at the sample 
boundary with opposite spins, often dubbed as {\it helical} edge modes
\cite{Andrei}.
The idea of "$Z_2$" topological insulator stems from the observation that 
these helical edge modes are robust against weak perturbations, such as
the extrinsic Rashba spin-orbit interaction (coupling strength: $\lambda_R$).
Thus, non-trivial topological nature of a $Z_2$ topological insulator
is often attributed to the existence of such edge modes, protected by 
Kramers degeneracy.
This paper, on the contrary, highlights its {\it bulk} property.
Since real materials always have disorder, 
we investigate its transport property {\it under doping} using the framework of
standard weak localization theory.

Of course, the magnitude of spin-orbit interactions has always been
questioned in graphene
\cite{Min,Brataas,SCZ},
leading to search for $Z_2$ nature in a system of larger spin-orbit coupling
\cite{Andrei,Laurens}.
The existence of helical edge modes was first experimentally
shown in a two-dimensional HgTe/CdTe heterostructure \cite{Laurens}.
Recall that in graphene
two doubly degenerate Dirac cones appear at $K$- and $K'$- points
in the first Brillouin zone \cite{Geim}, in contrast to 
a single pair of Dirac cones appearing at the $\Gamma$-point in HgTe/CdTe qauntum well.
The first estimate of $\Delta$ and $\lambda_R$ in the original paper of Kane and Mele:
$2\Delta\sim$2.4 K, and
$\lambda_R/2\sim$ 0.5 mK for a typical strength of perpendicular electric field 
$E=$ 50 V/300 nm,
provides a favorable condition for $Z_2$ non-trivial phase
\cite{KM1}.
This estimate was later shown to be too optimistic (for the occurrence of $Z_2$ phase)
due to the specific geometry of $s$ and $p$ orbitals in graphene.
According to Refs. \cite{Min},
the estimated value of $\Delta$ ($\lambda_R$) is much smaller (larger) than the original 
estimation of Ref. \cite{KM1}:
$2\Delta\sim$0.01 K, and $\lambda_R/2\sim$ 0.13 K
for the same electric field of $E=$ 50 V/300 nm.
On the other hand, a recent first-principle calculation
suggests that $d$-orbitals play a dominant role in the gap opening
at $K$ and $K'$ points
\cite{d96}.
As a result, the actual value of $\Delta$ might be somewhat intermediate
between the previous estimates of \cite{KM1} and \cite{Min,Brataas,SCZ}, namely
$2\Delta\sim$0.28 K, $\lambda_R/2\sim$ 0.23 K per V/nm.
The concept of $Z_2$-topological insulator has also been extended to
three space dimensions
\cite{FK,SM,Joel,Qi,Sch,Hsieh1,RS,Hsieh2}.
A recent spin-ARPES study on Bi$_2$Te$_3$ reports on the experimental 
observation of a spin-helical two-dimensional surface state in such three-dimensional 
$Z_2$-topological insulator
\cite{Hsieh2}.

Localization properties of the doped Kane-Mele $Z_2$ insulator
have been studied numerically \cite{OBS, MO}.
Ref.\cite{MO} deduces a phase diagram in the $(E,W)$-plane ($E$: energy, $W$: strength of disorder),
in which a metallic domain appears in valence and conduction bands with a finite width in $E$.
As disorder is increased, these two extended domains in both bands approach to each other, and eventually merge and disappear.
A more subtle issue is the nature of the metallic state next to the $Z_2$ insulating phase.
It has been claimed \cite{MO} that 
the system's $Z_2$ symmetry leads to an unconventional symmetry class.
However, an extensive study on the critical exponents \cite{OBS}
has suggested that the weak anti-localization behavior of the doped $Z_2$ insulator belongs to the 
conventional symplectic symmetry class.
This paper addresses the basic mechanism 
how doped $Z_2$ insulators acquire such unique localization properties.
As a simple implementation of $Z_2$-topological insulator,
we consider Kane-Mele model, and in contrast to the numerical works of Refs.
\cite{OBS,MO}, 
we restrict our study to the limit of weak disorder.
On the other hand, we pay much attention to the existence of {\it valleys}
in graphene, since localization properties are much influenced by the presence 
or absence of scattering across different valleys in the Brillouin zone.
The later is determined by the range of the impurity potential
\cite{NA}.

This paper is organized as follows.
The Kane-Mele model is introduced in Sec. II.
Then, we apply the standard diagrammatic approach to weak localization
to the doped Kane-Mele model.
In Sec. III we consider the case of vanishing Rashba SOI.
Particular attention will be paid to different types of the mass term, (a) and (b),
together with the presence/absence of  $K$-$K'$ scattering.
Here, we will focus on unitary behaviors, which appear as a consequence of
a finite lifetime acquired by Cooperons.
Breaking or preserved effective time reversal symmetry will be the main issue
of this section.
Sec. IV is devoted to study on the effects of Rashba spin-orbit interaction.
In the final section, we will summarize our results, and  
give interpretation to them in terms of the number of {\it active} species of effective spins
\cite{IKN}.

\begin{table*}[htdp]
\caption{
Three strory structure of the Kand-Mele $Z_2$ topological insulator,
In addition to the real spin $\vec s$, there appear two types of pseudo-spins:
$\vec\sigma$ representing A-B sub-lattices, and $\vec\tau$ specifying the valley:
$K$ or $K'$. (*) If Fermi level is in the gap
}
\begin{center}
\begin{tabular}{c|c|c}
\hline\hline
&
long-range scatterers
&
short-range scatterers 
\\
&
(single valley)
&
($K$-$K'$ coupled by inter-valley scattering)
\\ \hline \hline
(i) ideal graphene
&
$H_1=\hbar v_F(p_x \sigma_x +p_y \sigma_y)$
&
$H_1=\hbar v_F(p_x \sigma_x \tau_z+p_y \sigma_y)$
\\
(massless)
&
single Dirac cone
&
opposite chiralities at $K$ and $K'$ 
\\ \hline
(ii) mass terms: (a) topological, (b) ionic
&
$H_2=m\sigma_z$
&
(a) $H_2= - \Delta\sigma_z\tau_z s_z \equiv H_\Delta$
\\
$\rightarrow$ (a) QSH, (b) ordinary insulator (*)
&
&
(b) $H_2=M\sigma_z \equiv H_M$
\\ \hline
(iii) Rashba spin-orbit interaction
&
\multicolumn{2}{c}{
$H_R= - \lambda_R(\sigma_x \tau_z s_y - \sigma_y s_x)$
}
\\
$\rightarrow$ $Z_2$ topological insulator
&
\multicolumn{2}{c}{
mixes real spin $\uparrow$ and $\downarrow$
(spin rotational-symmetry broken)
}
\\
\hline\hline
\end{tabular}
\end{center}
\label{default}
\end{table*}

\section{Kane-Mele model}

The Kane-Mele model is given a status as a prototype for 
various $Z_2$ topological insulator models.
It was introduced as a model for graphene in the presence of spin-orbit interactions
\cite{KM1}.
The model is first defined on the hexagonal lattice, in the framework of
tight-binding approximation.
The continuum limit is then taken, in which the effective Hamiltonian becomes,
\begin{eqnarray}
H_{KM}&=&H_1+H_\Delta+H_R
\nonumber \\
H_1&=&\hbar v_F(p_x \sigma_x \tau_z+p_y \sigma_y)
\nonumber \\
H_\Delta&=& -\Delta\sigma_z\tau_z s_z 
\nonumber \\
H_R&=& - {\lambda_R \over 2}(\sigma_x \tau_z s_y - \sigma_y s_x),
\label{HKM}
\end{eqnarray}
where three Pauli's matrices, $\vec{\sigma}$, $\vec{\tau}$, and $\vec{s}$
operate in different spaces. 
Namely, 
$\vec{\sigma}$ acts on pseudo-spin specifying the A-B sublattices, 
$\vec{\tau}$ on the $K$-$K'$ ``valley spin", 
and $\vec{s}$ on the real spin.
Throughout this paper, we assume that $\Delta >0$
\cite{Delta}.

\subsection{Three story structure}

The Kane-Mele model defined as Eq. (\ref{HKM}) has 
the following three story structure (see TABLE I), corresponding to each term 
of Eq.  (\ref{HKM}):
(i) graphene on its base,
(ii) a topological mass term, encoding quantized spin Hall (QSH) effect,
and finally
(iii) the Rashba spin-orbit interaction $\lambda_R$.
Let us first look into the role of these three floors one by one.

\subsubsection{Graphene and its localization properties}

Graphene, an isolated single layer of graphite, has a band structure, 
with two massless points (often referred to $K$ and $K'$) 
in the first Brillouin zone, and in the vicinity of these points the low-energy effective Hamiltonian 
reduces to a Dirac-Weyl form \cite{Geim}.
This part of the Hamiltonian, i.e., $H_1$ in Eq. (\ref{HKM}), 
comes from the standard nearest neighbor hopping term in the tight binding approximation.
Such ideal (massless) graphene
shows weak anti-locazation behavior \cite{SA,KN1,Bee}, when inter-valley ($K$-$K'$) 
scattering is irrelevant \cite{SA}.
The absence or presence of $K$-$K'$ scattering is determined by the range of 
scattering potential \cite{NA}.
%Here, we consider only scalar potential scatterers in contrast to
%conventional weak anti-localization theory
%that considers also magnetic and spin-orbit scatterings 
%\cite{HLN}.
Short- (long-) range scatterers do (not) see the difference between A and B sites, and also,
do (not) involve $K$-$K'$ scattering.
%Short-range scatterers, on the other hand, do distinguish A and B sites, involve
%$K$-$K'$ scattering.
The weak localization behavior of graphene is indeed %due to the presence
susceptible of presence or absence of $K$-$K'$ scattering
\cite{SA, Roche}.
In the absence of $K$-$K'$ scattering, it is now established \cite{Efetov}
that at one-loop order
the graphene (a single Dirac cone) shows a weak anti-localization behavior, 
indicating that the system is symplectic \cite{D}.

An interesting question is to what extent
this anti-localization tendency continues against generalizations.
%beyond one-loop order.
Recent numerical analyses suggest that this anti-localization tendency 
actually continues to the strong-coupling regime \cite{KN1,Bee},
due probably to some cancellation of higher order terms in the expansion of $\beta (g)$.
Ref. \cite{KN1} argues that
such cancellation of higher order terms
is a consequence of a non-trivial spectral flow, associated with $Z_2$
Kramers symmetry, from which they coined the word, ``$Z_2$ (topological) metal".
It is also pointed out that
unconventional behaviors of $Z_2$ metal can be also casted
in terms of a topological term in the effective $\sigma$-model description
\cite{Ryu}.
In this paper, we achieve another generalization of Ref.\cite{SA}
by reinterpreting the weak anti-localization property of graphene 
as the massless limit of a more general system, i.e.,
that of the Kane-Mele quantized spin Hall (QSH) insulator.

Let us go back to the explicit form of Eq. (\ref{HKM}), and focus on
its properties under time-reversal operation.
The time reversal symmetry (TRS) plays an essential role in the discussion of
symmetry classes in weak localization theory. 
In the absence of $K$-$K'$ scattering, the two Dirac cones are decoupled dynamically,
whereas, TRS operation transforms $K$ to $K'$, and 
vice versa.
In such cases, it is convenient to introduce the idea of 
``pseudo time reversal symmetry" (PTRS) \cite{Ludwig},
in which one pretends that $\vec{\sigma}$ represents a real spin 
so that pseudo-Kramers' pairs are formed in a single Dirac cone.
Then PTRS transforms $\vec{p}$ to $-\vec{p}$ within a valley.
Since $\vec{\tau}$ is invariant under PTRS by definition,
$H_1$ is also invariant.
Unlike the real spin, however, the pseudo-spin should have single-valued eigenstates.  
This apparent difficulty is resolved in terms of the Berry phase \cite{NA}
as discussed shortly toward the end of this section. 

\subsubsection{Topological mass and the quantized spin Hall effect}

In the second row of TABLE I shows,
the two Dirac cones at $K$ and $K'$ valleys of graphene acquire a gap 
(the total Hamiltonian becomes $H=H_1+H_2$)
either in the presence of 
(a) imaginary hopping between second-nearest neighbors due to 
spin-orbit interaction \cite{KM1}, or
(b) AB sublattice symmetry breaking staggered chemical potential.
In case (a), the so-called topological mass term $H_\Delta$ is generated,
whereas (b) leads to a standard ionic mass term $H_M$.

First note that in case (a),
the topological mass term $H_\Delta= - \Delta\sigma_z\tau_z s_z$ 
is time reversal invariant.
This stems from the fact that spin-orbit interaction preserves TRS.
Second, note also that this is true 
only when we count both the (real) spin up and down
sectors.
Namely, if we pick up, say, only up spin part, then the system
breaks TRS, showing, e.g., a finite $\sigma_{xy}$ ($=\pm e^2/h$, see below) 
even in the absence of magnetic field \cite{HAL'88}.
On the other hand, (b) a staggered chemical potential generates a mass term of the form:
$H_M=M\sigma_z$, which has
the same sign at $K$ and $K'$ points.
The total Hamiltonian $H=H_1+H_M$ describes an ordinary insulator such as a monolayer 
of boron nitride (BN).

Consideration on such different types of energy gap \cite{HAL'88} naturally
leads to the idea of ``quantized" spin Hall insulator \cite{KM1}.
In contrast to the ordinary ionic mass case (b), where
the so-called parity anomaly cancels between the $K$- and $K'$-points 
and does not manifest itself \cite{SEM}, 
the topological mass term (a) 
has an opposite sign at $K$- and $K'$-points \cite{HAL'88}
as well as for up and down spins, 
resulting in the quantized spin Hall effect. 
There are actually two copies of 
quantum Hall states \`a la Ref. \cite{HAL'88} under zero magnetic field,
one with spin $\uparrow$ and $\sigma^\uparrow_{xy}=+e^2/h$, 
and the other with spin $\downarrow$ and $\sigma^\downarrow_{xy}=-e^2/h$
\cite{KM1}.

The existence of an energy gap is also suggested in experiments.
Its magnitude is under debate in photoemission experiments \cite{Bostwick,DHLee}
In contrast to the theoretical prediction
\cite{SA, Efetov, KN1},
weak localization experiments on graphene \cite{Morozov,Gorby,SiC} show also a unitary behavior.
It is, therefore, natural to ask how the localization properties would be 
influenced by the presence of mass term.
Absence of WL may be attributed to ripples \cite{Guinea,KN2}.

The type of mass term, given either by 
$H_\Delta$ or $H_M$, is a relevant factor in our discussion on localization properties
(see later sections).
If there is no $K$-$K'$ scattering, and two Dirac cones are decoupled,
however, the system cannot see the difference between the two types of mass term.
The behavior of the system is thus strongly dependent on the 
presence or absence of $K$-$K'$ scattering. 
We will see in later sections,
when $K$-$K'$ scattering is switched on, the topological mass (b) leaves the system unitary,
whereas the ordinary mass drives the system to orthogonal symmetry class.
Our analysis will be summarized in the language of PTRS  in the final section
(see TABLE II), 
and  should be applicable to
%here may also be regarded as weak (anti-)localization study on a toy model of 
%more realistic but certainly more complicated, and sometimes higher dimensional
two dimiensional $Z_2$ insulators in general.

%We have already argued that as far as $\lambda_R=0$,
%the Kane-Mele system can actually be viewed as 
%{\it two copies of decoupled quantum Hall states},
%the latter characterized by the so-called U(1) Chern numbers \cite{U(1)}.
%%indicating that the system's underlying symmetry is U(1).
%The system has also U(1) spin rotational symmetry
%around the spin quantization axis ($s_z$-axis in our notation),
%since the Hamiltonian $H_2$ in TABLE I has no off-diagonal term (spin-flip)
%in real spin space such as $s_x$ or $s_y$.
%The full SU(2) spin rotational symmetry is thus lost by the topological mass term
%%in our system 
%(even without Rashba term).
%The spin quantization axis $s_z$ can be taken arbitrarily, i.e., 
%independent of the real space coordinates \cite{HLN+}.

\subsubsection{Rashba spin-orbit interaction and $Z_2$ topological order}

Let us finally consider the third row (iii) of TABLE I, in which the total Hamiltonian becomes
$H=H_1+H_2+ H_R$.
The Rashba term $H_R$ turns out to be a relevant perturbation to the 
above symmetry properties.
First, as for topological properties of the undoped phase, 
the quantized spin Hall effect is not robust, but replaced by a $Z_2$ topological order.
\cite{KM2}

As for weak localization properties of the doped phase,
the Rashba term $H_R$, fixing the relative angle between $s_z$ and
the real space coordinates, changes the symmetry class.
We will see in later sections that
the Rashba spin-orbit interaction turns the system from
unitary to orthogonal in the absence of $K$-$K'$ scattering, 
whereas in the presence of $K$-$K'$ scattering, the system turns from
unitary to symplectic with weak anti-localization behavior.
%Certainly, identification of the universality class is much beyond the scope of our weak localization 
%study, but still the 
In the present ``poor man's" analysis,
% suggests 
we can identify the scattering channel that separates these different symmetry classes.
The two weak anti-localization phases, 
one in the graphene limit (single Dirac cone, unconventional)
and the other associated with a $Z_2$ topological insulator, 
%are {\it not smoothly connected}.
evolve from each other via either orthogonal or unitary behavior, activated by the Rashba term together with $K$-$K'$ scattering or the topological mass, respectively.

\vspace{0.5 cm}
Let us emphasize that the weak anti-localization behavior of graphene occurs in the phase diagram of 
{no $K$-$K'$ scattering}, wheras the weak anti-localization of $Z_2$ topological insulator occurs 
{\it due to $K$-$K'$ scattering}.
In the previous numerical analysis \cite{MO,OBS}, this important fact has not been noticed because the disorder in the real-space model always involved the effective inter-valley scattering.

\subsection{Construction of eigenstates}

To construct eigenstates explicitly,
we first consider the simplest non-trivial case with vanishing Rashba interaction,
and allow only long-range scatterers.
Since $\lambda_R =0$, real spin up and down sectors become decoupled.
In terms of the Hamiltonian, $H=H_1+H_2$ is diagonal
in real spin $\vec{s}$-space as well as in $\tau$-spin space.
We can, therefore, consider separately $\tau_z=1$ ($K$-valley) and $\tau_z=-1$ ($K'$-valley).
Taking $\hbar=v_F=1$ for simplicity, 
one may rewrite the Hamiltonian in the $K$-valley, $H=H_1+H_\Delta$
(with $\tau_z=1$, $s_z=1$), in the following simple form:
\begin{equation}
H=\vec{p}\cdot \vec{\sigma},
\end{equation}
by introducing a fictitious three-dimensional momentum $\vec{p}=(p_x, p_y, -\Delta)$.
As for $H=H_1+H_M$, one has simply to replace it with
$\vec{p}=(p_x, p_y, M)$.

The Hamiltonian can then be diagonalized by choosing a proper 
quantization axis of the 
pseudo-spin, in analogy to the SU(2) spin case.
One must take into account here that the $\vec{\sigma}$ represents
only a pseudo spin, as it is derived from the sublattice degree of freedom. 
Correspondingly, the momentum $\vec{p}$, specifying the quantization axis for
$\vec{\sigma}$ is  {\it single-valued}.
Taking this single-valuedness also into account, one may denote the eigenvalues and
the eigenvectors of $H=H_1+H_2$ as
\begin{eqnarray}
H |\vec{p}\pm\rangle=\pm |\vec{p}|  |\vec{p}\pm\rangle
=\pm\sqrt{p_x^2+p_y^2+\Delta^2}  |\vec{p}\pm\rangle,
\nonumber \\
|\vec{p}+\rangle=
\left(
\begin{array}{c}
 \cos{\theta\over 2} \\
e^{i\phi}\sin{\theta\over 2}
\end{array}
\right),
|\vec{p}-\rangle=
\left(
\begin{array}{c}
 \sin{\theta\over 2} \\
- e^{i\phi} \cos{\theta\over 2}
\end{array}
\right),
\label{pseudo}
\end{eqnarray}
where $\theta$ and $\phi$ are polar angles
satifying
\begin{eqnarray}
\cos\theta={-\Delta\over \sqrt{p_x^2+p_y^2+\Delta^2}},\
%\cos\theta={\Delta\over \sqrt{p_x^2+p_y^2+\Delta^2}},\
%because I changed the sign of \Delta
\cos\phi={p_x\over \sqrt{p_x^2+p_y^2}}.
\end{eqnarray}
Here $|\vec{p}\pm\rangle$ corresponds  to the upper- (lower-) band eigenvector.
%Note that the pseudospin eigenstates (\ref{pseudo}) are single-valued. 
%aleady said ten lines before imura
In the course of an adiabatic evolution of 
$|\vec{p}(t)\pm \rangle$ around the origin of $\vec{p}$, however,
a Berry phase $\pi$ enters per winding.
\cite{Berry}
This situation keeps consistency with the double-valued SU(2) eigenstates of a 
real spin \cite{ANS}. 

In the following we will focus on 
the upper (conduction) band with $E=|\vec{p}| \ge \Delta$.
One can also write down the eigenstates
in the $K'$-valley using the same parametrization.
Because $\tau_z=-1$ in the $K'$-valley, the conduction band has eigenfunctions of the form 
$|\vec{p}-\rangle$ in Eq.(\ref{pseudo}) with $\phi$ replaced by $-\phi$.
We also introduce the notation, $\alpha$, $\alpha'$, $\beta$, etc.
to specify the momentum $\vec{p}$,
in order to keep the consistency in notation with Ref.\cite{SA}.
For example, we will use notations such as,
\begin{eqnarray}
|K\alpha\rangle&=&|\vec{p}+\rangle=
\left(
\begin{array}{c}
 \cos{\theta_\alpha\over 2} \\
e^{i\phi_\alpha}\sin{\theta_\alpha\over 2}
\end{array}
\right),
\label{ketK} \\
|K'\alpha\rangle&=&
\left(
\begin{array}{c}
e^{i\phi_\alpha}\sin{\theta_\alpha\over 2} \\
- \cos{\theta_\alpha\over 2}
\end{array}
\right).
\label{ketK'}
\end{eqnarray}
As long as the Rashba term is absent,
one can safely fix $\vec{s}$, say, to be $\uparrow$.
Then we do not explicitly consider the real spin until Section \ref{Rashba}.

\section{Weak localization property --- unitary cases}

\begin{figure}[htdp]
\includegraphics[width=7.5 cm]{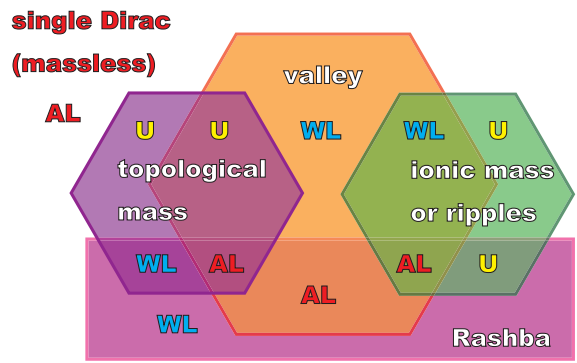}
\caption{(Color online) Weak localization phase diagram of the doped Kane-Mele model
in the presence of potential scatterers.
Relation to other graphene-based models with different types of the mass, or ripples
is taken into account.
WL, AL and U refer to weak localization (orthogonal class), 
weak anti-localization (symplectic class) and absence of WL
(unitary class), respectively.}
\label{venn}
\end{figure}

Weak localization (WL) phenomena have been known since three decades
\cite{CSB}.
Scaling to metal (weak anti-localization) was shown to be possible
due to scattering by spin-orbit interaction \cite{HLN}.
Absence of WL (unitary behavior) is, on the other hand, attributed to explicit breaking of
time reversal symmetry (TRS).
This paper shows that, in systems of graphene,
a zoo of such different localization behaviors 
appears under the same Hamiltonian, simply by activating or inactivating
effective spin degrees of freedom (FIG. \ref{venn}).
Graphene thus provides a contemporary aspect to the conventional WL 
theory framework.

%%%%%%%%%%%%%%%%%%%%%%%%%%%%%%%%%%%%%%%
%%%%%%%%%%%%%%%%%%%%%%%%%%%%%%%%%%%%%%%

We apply standard diagrammatic techniques for weak localization
to the doped Kane-Mele model.
In real systems, such as a graphene sheet, doping can be easily done
by simply applying a gate voltage.
Suppose that our Fermi level is in the conduction band, so that the system is 
metallic in the clean limit.
We then introduce weak disorder 
by taking into account scattering by impurities.
To characterize our two-dimensional system of size $L^2$,
we focus on its longitudinal conductivity $\sigma_{xx}=g(L)$.
Weak localization refers to $1/g$ correction to the so-called scaling function, 
\begin{equation}
\beta(g)={d \log g\over d \log L}=D-2-{c_1\over g}+\cdots,
\label{betag}
\end{equation}
where $\beta(g)\rightarrow 0$ ($g\rightarrow\infty$) in two spatial dimensions ($D=2$).
$1/g$-correction to $\beta (g)$ leads to logarithmic corrections to conductivity.
Thus, the sign of $1/g$-correction in Eq. (\ref{betag}), i.e., the sign of $c_1$ 
determines system's weak localization property:
(i) $c_1>0$: WL (orthogonal), (ii) $c_1=0$: absence of WL peak in resistance data
(unitary), and (iii) $c_1<0$ (symplectic).
Symmetry classes (orthogonal, unitary, and symplectic) are due to classification of
corresponding random matrices \cite{D}.
Such logarithmic corrections to conductivity can be calculated using
diagram techniques based on Kubo formula
\cite{AKLL}.
Diagrams contributing to a weak localization correction are
particle-particle type ladders, or
sometimes also called a ``Cooperon".
In contrast to the particle-hole type diagrams, which
{\it always} show a ``diffusion type" $1/q^2$ singularity,
whether Cooperon diagrams are susceptible of such $1/q^2$ singularity
is a more subtle issue related to time reversal symmetry of the system
\cite{CSB}.

In this section, we first switch {\it off} Rashba spin-orbit interaction, and study
whether different types of mass term lead to the absence of WL peak
(unitary behavior).
We consider only scalar potential scatterers in contrast to Ref. \cite{HLN}.
Instead, we distinguish impurities of different potential range, and
classify them into two categories, depending on whether they involve
inter-valley scattering or not.
Long (short)-range scatterers involve (do not involve) $K$-$K'$ scattering.

\subsection{Long-range scatterers}
% --- crossover from WAL to WL}
Long-range scatterers involve only intra-valley scattering.
%(the impurity vertex is (proportional to) 1 in the $K$-$K'$ space).
One can, therefore, safely focus on, say, the $K$-valley.
Such scatterers do not distinguish between A and B sublattices, either, i.e.,
the impurity vertex is also unity in the $AB$ sublattice space.
The matrix element associated with long-range scatterers is, therefore,
proportional to, 
\begin{equation}
\langle K\beta|1|K\alpha\rangle=\cos^2{\theta\over 2}+e^{i(\phi_\alpha-\phi_\beta)}
\sin^2{\theta\over 2}
\label{phase}
\end{equation}
where we assumed elastic scattering ($\theta_\alpha=\theta_\beta=\theta$).
Note that here we pick up only the spin part of the matrix element,
and $|K\alpha\rangle$ represents only the spinor part of the electron wave function.
In order to find the full matrix element between different momentum eigenstates,
Eq.(\ref{phase}) should be appended by 
%compensated 
by a spatial part that requires the momentum conservation.
The phase factor in Eq. (\ref{phase}) is analogous to the Berry phase,
which has already appeared in Ref. \cite{ANS}, in the graphene limit
($\theta\rightarrow \pi/2$).
The Berry phase, in the presence of a mass term, is not generally $\pi$
in contrast to the massless limit.
Note also that Eq.(\ref{phase}) involves an imaginary number
that indeed turns the system from weak localization to weak anti-localization.
In the present case, its complex nature 
does not come from the scattering potential, but
from the property of wave function.
%This is quite contrasting to more conventional more familiar studies on weak anti-localization systems \cite{HLN}.

From the self-energy diagram, we define the scattering time $\tau_L$:
\begin{eqnarray}
{1\over \tau_L}&=&2\pi\nu (E) n_L u_L^2
\left\langle |\langle K\beta|1|K\alpha\rangle|^2\right\rangle
\nonumber \\
&=& \eta_L
%2\pi\nu n_L u_L^2 a
\left(\cos^4{\theta\over 2}+\sin^4{\theta\over 2}\right),
\label{tauL}
\end{eqnarray}
where $n_L$, $u_L$ and $\nu (E)$ are, respectively, the impurity density, the strength of impurity scattering potential, and the density of states at the given energy $E$.
They combine to give $\eta_L= 2\pi\nu (E)n_L u_L^2$, and
$\left\langle \cdots\right\rangle$ represents the angular part of
impurity average.

%\vspace{0.3cm}
%\noindent
%{\it Remark} --- In contrast, 
The transport relaxation time $\tau_{tr}$ 
%(with vertex correction) is susceptible of 
involves the $\cos(\phi_\alpha-\phi_\beta)$ term as
%\begin{equation}
%|\langle K\beta|1|K\alpha\rangle|^2=
%\cos^4{\theta\over 2}+\cos(\phi_\alpha-\phi_\beta){\sin^2 \theta \over 2}
%+\sin^4{\theta\over 2}.
%\nonumber
%\end{equation}
\begin{eqnarray}
{1\over\tau_{tr}}&=& \eta_L
\left\langle |\langle K\beta|1|K\alpha\rangle|^2
[1-\cos(\phi_\alpha-\phi_\beta)]\right\rangle
\nonumber \\
&=&\ \eta_L \left[\cos^4{\theta\over 2}+\sin^4{\theta\over 2}
-{\sin^2 \theta \over 4}
\right]
\label{tautr}
\end{eqnarray}
where we have used 
$\langle \cos(\phi_\alpha-\phi_\beta) \rangle=0$, 
$\langle \cos^2(\phi_\alpha-\phi_\beta) \rangle=1/2$.
Usually, 
the factor $1-\cos(\phi_\alpha-\phi_\beta)$ in 
the first line of Eq.(\ref{tautr})
is inactive for $\delta$-function like ($s$-wave) scatterers. 
Here, the matrix (Dirac)
nature of the Hamiltonian induces a cosine term that corresponds to
vertex correction in diagrammatic language.

\begin{figure}[!]
\includegraphics[width=7.5 cm]{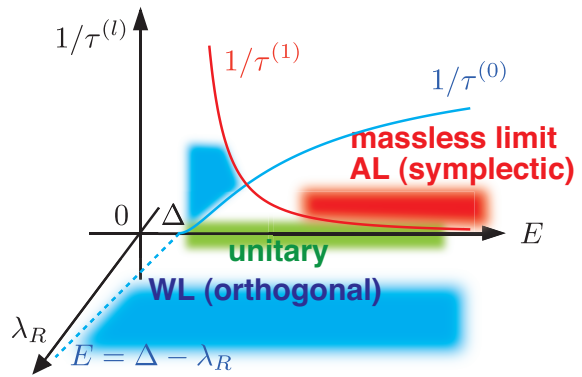}
\caption{(Color online) Weak localization properties 
for long-range scatterers (LRS) without $K$-$K'$ scattering.
The ordinate shows
crossover from weak anti-localization (WAL) to weak localization (WL) tendency
($\lambda_R=0$, as $E$ is decreased from the graphene limit toward the bottom of the band).
Crossover from unitary to orthogonal symmetry class ($\lambda_R \ne 0$).}
\label{LRS}
\end{figure}

The bare vertex function $\gamma$ has also $\phi$-dependence:
\begin{eqnarray}
\gamma&=& \eta_L\left[
\cos^4{\theta\over 2}+e^{i(\phi_\alpha-\phi_\beta)}{\sin^2 \theta\over 2}
+e^{2i(\phi_\alpha-\phi_\beta)}\sin^4{\theta\over 2}
\right]
\nonumber \\
&=& \gamma^{(0)}+\gamma^{(1)} e^{i(\phi_\alpha-\phi_\beta)}
+\gamma^{(2)} e^{2i(\phi_\alpha-\phi_\beta)}.
\label{gamma012}
\end{eqnarray}
In the second line,
we classify terms 
according to their relative angular momentum, i.e., different $\phi$-dependence:
$e^{il(\phi_\alpha-\phi_\beta)}$, where $l=0,1,2$.
This expansion helps to solve the Bethe-Salpeter equation:
\begin{equation}
\Gamma_{\alpha\beta}=\gamma_{\alpha\beta}+\gamma_{\alpha\mu}\Pi_\mu\Gamma_{\mu\beta}.
\label{BSE}
\end{equation}
%The Bethe-Salpeter equation gives the diagnosis for constructing the full dressed vertex function $\Gamma$.
To find the solution of Eq. (\ref{BSE}), 
we expand also $\Gamma$ into different angular momentum components:
\begin{equation}
\Gamma_{\alpha\beta}=\Gamma^{(0)}+\Gamma^{(1)} e^{i(\phi_\alpha-\phi_\beta)}+
\Gamma^{(2)} e^{2i(\phi_\alpha-\phi_\beta)},
\end{equation}
and integrate over $\phi_\mu$, i.e., over the intermediate angle dependence in Bethe-Salpeter equation.
Notice also that $\Pi\simeq\tau_L(1-\tau_L Dq^2)$, with $D$ being the diffusion constant: 
$D=v_F^2\tau_L/2$, 
one finds, at the same order of precision,
\begin{equation}
\left[1-\gamma^{(l)} \tau_L(1-\tau_L Dq^2)\right]\Gamma^{(l)}=\gamma^{(l)}.
\label{BSE_l}
\end{equation}
The crucial issue is the 
cancellation (or not) of the leading order ($\sim 1$) term in the
coefficient of $\Gamma^{(l)}$.
In the graphene limit 
($\theta=\pi/2$), 
$\Gamma^{(1)}$ becomes $\sim 1/q^2$-singular
driving the system to weak anti-localization.
At the bottom of conduction band, on the other hand, 
$\Gamma^{(0)}$ becomes more important, and
toward the limit $\theta\rightarrow 0$ (though the model becomes ill-defined in this limit),
it tends to show $\sim 1/q^2$-singularity, 
leading the system to 
weak localization.
Away from these limits, 
the system shows a unitary behavior, i.e.,
all the three Cooperons acquire a finite lifetime.

Here, let us recall that the weak localization refers
to a logarithmic correction to the longitudinal conductivity $\sigma_{xx}$,
of the form, 
\begin{equation}
\Delta \sigma_{xx} \sim \mp A \log {L \over \bar{l}}
\label{logdiv}
\end{equation} 
($L$: size of the system, $\bar{l}$: mean free path, $A$: constant of order $e^2/h$).
In front of the logarithmic term,
$-$ sign should be chosen for the weak localization case.
Whereas, in the case of weak anti-localization, this overall sign in front of the logarithmic divergence 
is positive (the correction tends to increase the conductivity).
In the unitary case we mentioned above,
the logarithmic divergence of Eq. (\ref{logdiv}) as $L\rightarrow \infty$
is cut off by the longest lifetime of a Cooperon.
And in that sense,
the behavior of the system is driven by the Cooperon of the longest lifetime.
This situation is analogous to the case of spin-dependent scattering
studied in Refs. \cite{HLN}.
In order to quantify such unitary behaviors,
we define the lifetime  $\tau^{(l)}$ of a Cooperon, $\Gamma^{(l)}$ such that
\begin{equation}
\Gamma^{(l)}=
\frac 1 
{\tau_L^2
(Dq^2+1/\tau^{(l)})
},
\end{equation}
where all $\tau^{(l)}$'s are written in terms of a single parameter
$\tan \theta/2$:
\begin{eqnarray}
{\tau_L \over \tau^{(0)}} &=& \tan^4 {\theta \over 2},
\nonumber \\
{\tau_L \over \tau^{(1)}} &=& {1\over 2} \left({1\over \tan (\theta/2)}-\tan {\theta \over 2}\right)^2,
\nonumber \\
{\tau_L \over \tau^{(2)}} &=& {1 \over\tan^4 (\theta/2)}.
\end{eqnarray}
Since we have $\tan (\theta/2) \leq 1$, there is no chance for $\Gamma_2$ to dominate.
The parameter $\theta$ is transformed to energy $E=\Delta/\cos\theta$. As $E$ increases from $E=\Delta$, 
which means that $\theta$ increases from $\theta=0$, 
a crossover occurs at  
``universal" value,
\begin{equation}
\theta=\theta_c=2 \arctan {1\over\sqrt{2}}= 1.23\cdots.
\end{equation}
When $\theta_c<\theta<\pi/2$, $\Gamma^{(1)}$ (weak anti-localization) becomes dominant ,
leading to a {\it positive} logarithmic correction to the longitudinal conductivity,
of the form of Eq. (\ref{logdiv}) with  the $+$ sign, but $L$ replaced by $\tau^{(1)}$.
When $\theta<\theta_c$, $\Gamma^{(0)}$ (weak localization) is dominant,
and the correction to $\sigma_{xx}$ is given by Eq. (\ref{logdiv}) with the $-$ sign.
In terms of $E$, the crossover occurs at $E=E_c=3\Delta$, and at this point
the logarithmic correction changes its sign.
This crossover behavior is illustrated in FIG. 1, where 
we consider $\lambda_R=0$ for the moment.

Since long-range scatterers do not involve $K$-$K'$ scattering, and two Dirac cones are decoupled,
the system cannot see the difference between the two types of mass terms.
Thus for both types of mass terms, we found predominantly a unitary behavior.
Cooperons acquire a finite lifetime, which plays the role of cutting off 
the logarithmic correction, Eq. (\ref{logdiv}).
Typically, there is no $1/g$-correction to the $\beta (g)$-function,
in the $L\rightarrow \infty$ limit.
However, one can still see a crossover from anti-localization tendency to
weak localization regime,
as far as the longest lifetime of a Cooperon is much larger than the system size $L$.
When Fermi level is far above the gap, the electron does not feel very much that there is a gap,
i.e., he has a tendency to behave as if he were a massless Dirac fermion
(weak anti-localization behavior in the graphene limit, $E\rightarrow\infty$).
As the Fermi level approaches the bottom of the band, the electron
starts to feel the gap, and when he is close, he even forgets about that he is actually
a ``relativistic fermion", and starts to behave (say, $E<3\Delta$) as if he were a non-relativistic electron,
showing the weak localization behavior.

\subsection{Short-range scatterers with $K$-$K'$ scattering}
Short-range scatterers couple 
different valleys, i.e., $K$ and $K'$.
One may, therefore, possibly see the difference between two different types of masses; ordinary and topological.
The scattering matrix elements involve also a projection operator in the AB sublattice space,
i.e.,
\begin{equation}
{\cal P}_A=\left(
\begin{array}{cc}
1  &   0 \\
0  &   0
\end{array}
\right),\ \ \
{\cal P}_B=\left(
\begin{array}{cc}
0  &   0 \\
0  &   1  
\end{array}
\right).
\label{pApB}
\end{equation}
Matrix elements of such projection operators in the $K$-valley are,
\begin{eqnarray}
\langle K\beta|{\cal P}_A|K\alpha\rangle&=&\cos^2{\theta\over 2},
\nonumber \\
\langle K\beta|{\cal P}_B|K\alpha\rangle&=&e^{i(\phi_\alpha-\phi_\beta)}\sin^2{\theta\over 2}.
\label{SRSKK}
\end{eqnarray}
Matrix elements of such projection operators involving $K'$-valley depends,
on the contrary, on the type of mass.

\subsubsection{Topological mass case}
Let us first consider the case of Kane-Mele quantized spin Hall insulator, 
for which the conduction band eigenkets are given by Eqs. (\ref{ketK},\ref{ketK'}).
Since we are concerned about $K$-$K'$ scattering, 
let us first consider the inter-valley scattering matrix elements: %imura
\begin{eqnarray}
\langle K'\beta|{\cal P}_A \tau_-|K\alpha\rangle&=&
e^{-i\phi_\beta}\sin{\theta\over 2}\cos{\theta\over 2},
\nonumber \\
\langle K\beta'|{\cal P}_A \tau_+|K'\alpha'\rangle&=&
e^{i\phi_{\alpha'}}\sin{\theta\over 2}\cos{\theta\over 2},
\nonumber \\
\langle K'\beta|{\cal P}_B \tau_-|K\alpha\rangle&=&
- e^{i\phi_\beta}\sin{\theta\over 2}\cos{\theta\over 2},
\nonumber \\
\langle K\beta'|{\cal P}_B \tau_+|K'\alpha'\rangle&=&
- e^{-i\phi_{\alpha'}}\sin{\theta\over 2}\cos{\theta\over 2},
\label{SRSKK'}
\end{eqnarray}
where $\tau_\pm=(\tau_x \mp i\tau_y)/2$ are ``spin-flip" operators associated with 
the valley-spin ($K$-$K'$).
Their contribution to scattering time is, e.g.,
\begin{eqnarray}
&&2\pi\nu (E) n_A u_A^2
|\langle K'\beta|{\cal P}_A \tau_-|K\alpha\rangle|^2
\nonumber \\
&+&2\pi\nu (E) n_B u_B^2
|\langle K'\beta|{\cal P}_B \tau_-|K\alpha\rangle|^2
\nonumber \\
&=&2\gamma_S \sin^2{\theta\over 2}\cos^2{\theta\over 2},
\label{inter}
\end{eqnarray}
where we have defined
\begin{equation}
\eta_S=2\pi\nu (E) {n_A u_A^2+n_B u_B^2\over 2}.
\end{equation}
$n_{A,B}$ and $u_{A,B}$ are, respectively, the impurity density and the typical strength
of scattering potential at the A (B) sites.
In order to obtain the full expression for scattering time,
one has to consider also the contributions from intra-valley scattering, such as,
\begin{eqnarray}
2\pi\nu 
\left(
n_A u_A^2
|\langle K\beta|{\cal P}_A|K\alpha\rangle|^2
+n_B u_B^2
|\langle K\beta|{\cal P}_B|K\alpha\rangle|^2
\right)
\nonumber \\
=2\pi\nu 
\left(
n_A u_A^2 \cos^4{{\theta}\over 2}+
n_B u_B^2 \sin^4{{\theta}\over 2}
\right).
%\nonumber \\
%2\pi\nu
%\left(
%n_A u_A^2
%|\langle K'\beta|{\cal P}_A|K'\alpha\rangle|^2
%+n_B u_B^2
%|\langle K'\beta|{\cal P}_B|K'\alpha\rangle|^2
%\right)
%\nonumber \\
%=2\pi\nu (E) 
%\left(
%n_A u_A^2 \sin^4{{\theta}\over 2}+
%n_B u_B^2 \cos^4{{\theta}\over 2}
%\right).
\label{intra}
\end{eqnarray}
Here, we assumed that the strength of inter-valley scattering is the same
as intra-valley scattering, but this simplification is irrelevant to our conclusions. 
The scattering time in the $K$-valley reads,
\begin{eqnarray}
{1\over \tau_K}=
2\pi\nu n_A u_A^2
\left(
|\langle K\beta|{\cal P}_A|K\alpha\rangle|^2
+|\langle K'\beta|{\cal P}_A \tau_-|K\alpha\rangle|^2
\right)
\nonumber \\
+2\pi\nu n_B u_B^2
\left(
|\langle K\beta|{\cal P}_B|K\alpha\rangle|^2
+|\langle K'\beta|{\cal P}_B \tau_-|K\alpha\rangle|^2
\right),
\label{tauK}
\end{eqnarray}
and one finds a similar expression for $\tau_{K'}$.
One can verify, using Eqs. (\ref{inter},\ref{intra}), 
that as far as
$n_A u_A^2=n_B u_B^2\equiv n_S u_S^2$,
one finds,
\begin{eqnarray}
1/\tau_K=1/\tau_{K'}=\eta_S\equiv 1/\tau_S.
\end{eqnarray}
For short-range scatterers we consider here, the transport relaxation time is identical to
$\tau_S$, since
the projection in the AB sublattice space leaves
no cross term, i.e., $\phi$-dependent term; typically,  $\sim\cos(\phi_\alpha-\phi_\beta)$,
in the expression for $1/\tau_K$ and $1/\tau_{K'}$. 

\begin{figure}[!]
\includegraphics[width=7.5 cm]{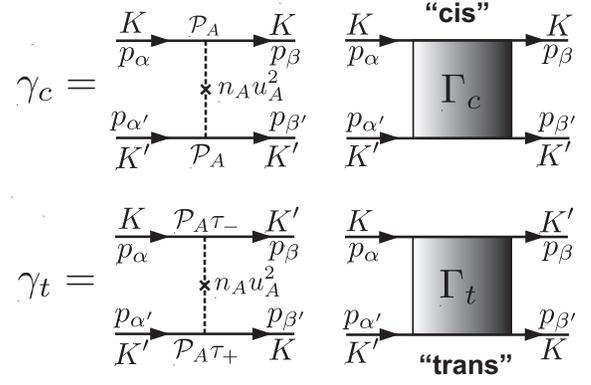}
\caption{Particle-particle ladders. Bare and dressed Cooperons. 
Relevant diagrams in the $KK'$ sector. 
``cis" and ``trans" refers to specific configurations of the valleys: $K$ and $K'$.}
\label{ladder_ct}
\end{figure}

As for particle-particle ladders, the momentum conservation naturally leads us to classify
them into $KK$, $KK'$-mixed, and $K'K'$ sectors.
They correspond in the notation of Ref.\cite{SA}, respectively, to 
$J=2, 0$ and $-2$ sectors. 
$J=j_\alpha+j_{\alpha'}=j_\beta+j_{\beta'}$ is conserved, 
where $j_\alpha=\pm 1$ if $\alpha$ occurs in the $K$- ($K'$-) valley,
since $\vec{K}-\vec{K'}$ is only half of a reciprocal lattice vector.

In the graphene limit, \cite{SA} the $KK'$-mixed sector is most divergent.
In the $J=0$ sector, two types of Cooperon diagrams are possible
(see FIG. 2).
Both of them have two $K$ electron and two $K'$ electron lines,
but they appear either in the ``cis" or ``trans" arrangement
(in the terminology of organic chemistry).
Naturally, $\gamma_c$ ($\gamma_t$) refers to cis (trans), and the same rule applies to
$\Gamma_{c,t}$.
The key issue here is that as a result of projection ${\cal P}_{A,B}$ and $K$-$K'$ scattering,
$\gamma_{t}$ acquires an additional minus sign:
\begin{eqnarray}
\gamma_{c} &=&
2\pi\nu n_A u_A^2
\langle K\beta|{\cal P}_A|K\alpha\rangle
\langle K\beta'|{\cal P}_A|K\alpha'\rangle
\nonumber \\
&+&2\pi\nu n_B u_B^2
\langle K'\beta|{\cal P}_B|K'\alpha\rangle
\langle K'\beta'|{\cal P}_B|K'\alpha'\rangle
\nonumber \\
&=& e^{i(\phi_\alpha-\phi_\beta)}
\eta_S {\sin^2 \theta\over 2}
\nonumber \\
&\equiv& \gamma_{c}^{(1)} e^{i(\phi_\alpha-\phi_\beta)}
\nonumber \\
\gamma_{t} &=&
2\pi\nu n_A u_A^2
\langle K'\beta|{\cal P}_A \tau_-|K\alpha\rangle
\langle K\beta'|{\cal P}_A \tau_+|K'\alpha'\rangle
\nonumber \\
&+&2\pi\nu n_B u_B^2
\langle K'\beta|{\cal P}_B \tau_-|K\alpha\rangle
\langle K\beta'|{\cal P}_B \tau_+|K'\alpha'\rangle
\nonumber \\
&=& - e^{i(\phi_\alpha-\phi_\beta)}
\eta_S {\sin^2 \theta\over 2}
\nonumber \\
&\equiv& \gamma_{t}^{(1)} e^{i(\phi_\alpha-\phi_\beta)}
\ (=-\gamma_{c}),
\label{gamma_ct}
\end{eqnarray}
canceling with the Berry phase.
Note also that as for $\phi$-dependence
both $\gamma_c$ and $\gamma_t$ have
only the $l=1$ component.
%$\gamma_{c,t}=\gamma_{c,t}^{(1)} e^{i(\phi_\alpha-\phi_\beta)}$.
Eq. (\ref{gamma_ct}) is a simple consequence of the matrix elements in Eq. (\ref{SRSKK'}).
Recall also that $\phi_\alpha-\phi_{\alpha'}=\pi$, since
$k_\alpha+k_{\alpha'}=k_\beta+k_{\beta'}=q \simeq 0$.

In order to calculate the correction to conductivity,
we set $\beta=\alpha'$, meaning $k_\alpha+k_\beta=q\simeq 0$,
therefore, $\phi_\alpha-\phi_\beta=\pi$.
%Though only $\Gamma_{t}$ gives a correction to conductivity, i.e., to the $\beta$-function,
The Bethe-Salpeter equation becomes two coupled equations:
\begin{equation}
\left(
\begin{array}{c}
\Gamma_{c}\\
\Gamma_{t}
\end{array}
\right)_{\alpha\beta}
=\left(
\begin{array}{c}
\gamma_{c}\\
\gamma_{t}
\end{array}
\right)_{\alpha\beta}
+\left(
\begin{array}{cc}
\gamma_{c}&\gamma_{t}\\
\gamma_{t}&\gamma_{c}
\end{array}
\right)_{\alpha\mu}
\Pi_\mu
\left(
\begin{array}{c}
\Gamma_{c}\\
\Gamma_{t}
\end{array}
\right)_{\mu\beta}
\label{BSE_ct}
\end{equation}
Indeed, both $\Gamma_{c}$ and $\Gamma_{t}$ contribute to the $1/q^2$ singularity.
After diagonalization, one finds,
\begin{eqnarray}
\Gamma_{c} +\Gamma_{t}&=&0,
\nonumber \\
\left[1-(\gamma^{(1)}_c -\gamma^{(1)}_t)\Pi_S \right](\Gamma_{c} -\Gamma_{t})&=&\gamma_{c} -\gamma_{t},
\label{BSE_1}
\end{eqnarray}
where  $\Pi_S\simeq \tau_S(1-\tau_S Dq^2)$.
The cancellation 
of the leading order ($\sim 1$) term 
is incomplete, giving a finite life time $\tau_{KK'}$ to the Cooperon:
\begin{equation}
\Gamma_{t} = {- e^{i(\phi_\alpha-\phi_\beta)}
\over 2\tau_S^2(Dq^2+1/\tau_{KK'})}.
\label{add-}
\end{equation}
The lifetime $\tau_{KK'}$ behaves, as a function of $E$, like
\begin{equation}
{\tau_S\over \tau_{KK'}}= \cot^2 \theta= {\Delta^2\over E^2-\Delta^2}.
\end{equation}
Clearly, this $KK'$ ($J=0$) Cooperon mode shows $1/q^2$-singularity only at the 
$E\rightarrow \infty$ limit \cite{SA}.
Decreasing energy toward the bottom of the band, another KK ($J=2$)-mode
becomes important.
The lifetime of the latter Cooperon (KK-mode) behaves, as a function of $E$,
like 
\begin{equation}
{\tau_S\over \tau_{KK}}= {E^2-\Delta^2 \over E^2+\Delta^2}.
\end{equation}
Thus, except at $E=\Delta$ and $E\rightarrow \infty$,
all the Cooperons having a finite life time, the system shows a unitary behavior
with no $1/g$-correction in the $L\rightarrow \infty$ limit (see FIG. 3).

One may find this contradictory to the fact that
the topological mass term $-\Delta\sigma_z\tau_z s_z$ does preserve TRS. 
To clarify this point, first note that the entire Hamiltonian, Eq.(\ref{HKM}) is time reversal
invariant, and that this comes from the fact that spin-orbit interaction preserves TRS.
Second, note also that this is true {\it only when we count both the (real) spin up and down
sectors}, and that if we pick up, say, only up spin part, then the system, showing, e.g.,
a finite $\sigma_{xy}$ ($=\pm e^2/h$) even in the absence of magnetic field \cite{HAL'88},
clearly breaks TRS.
In the absence Rashba spin-orbit interaction, spin up and down sectors are actually 
decoupled, indicating that the system belongs to the unitary class.

\begin{figure}[!]
\includegraphics[width=7.5 cm]{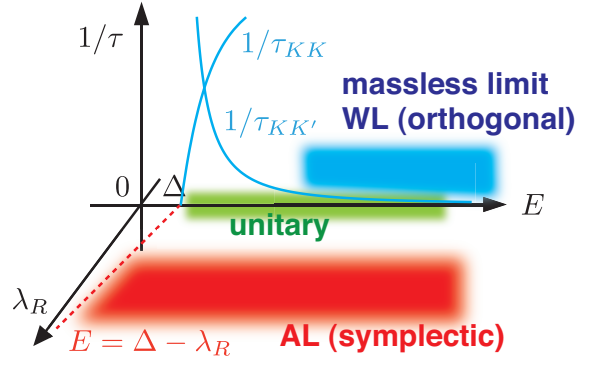}
\caption{(Color online) Topological mass case. 
Weak localization properties in the presence of short-range scatterers:
inter-valley scattering is allowed ($K$-$K'$ coupled).}
\label{SRS}
\end{figure}

\subsubsection{Ordinary mass case}
In the case of ordinary mass, one has to replace
$|K'\alpha\rangle$ in Eq.(\ref{ketK'}) with
\begin{equation}
|K'\alpha\rangle=
\left(
\begin{array}{c}
e^{i\phi_\alpha}\cos{\theta_\alpha\over 2} \\
-\sin{\theta_\alpha\over 2}
\end{array}
\right).
\end{equation}
In this basis, 
inter-valley matrix elements become
\begin{eqnarray}
\langle K'\beta|{\cal P}_A \tau_-|K\alpha\rangle&=&
e^{-i\phi_\beta}\cos^2{\theta\over 2},
\nonumber \\
\langle K\beta'|{\cal P}_A \tau_+|K'\alpha'\rangle&=&
e^{i\phi_{\alpha'}}\cos^2{\theta\over 2},
\nonumber \\
\langle K'\beta|{\cal P}_B \tau_-|K\alpha\rangle&=&
- e^{i\phi_\beta}\sin^2{\theta\over 2},
\nonumber \\
\langle K\beta'|{\cal P}_B \tau_+|K'\alpha'\rangle&=&
- e^{-i\phi_{\alpha'}}\sin^2{\theta\over 2}.
\label{SRSKK'2}
\end{eqnarray}
Notice that their contribution to the scattering time is identical to Eq. (\ref{intra}).
One finds, therefore, substituting Eqs. (\ref{SRSKK}) and (\ref{SRSKK'2}) into Eq. (\ref{tauK}),
that the scattering time becomes this time,
\begin{eqnarray}
{1\over \tau_K}&=&4\pi\nu (E) 
\left(
n_A u_A^2 \cos^4{{\theta}\over 2}+
n_B u_B^2 \sin^4{{\theta}\over 2}
\right)
\nonumber \\
&=&2\eta_S
\left(
\cos^4{{\theta}\over 2}+
\sin^4{{\theta}\over 2}
\right)
\equiv {1\over \tau_S'}.
\label{tauS'}
\end{eqnarray}
One can also verify that $1/\tau_{K'}=1/\tau_K$.
On the other hand, the expressions for $\gamma_{c,t}$ become also,
\begin{eqnarray}
\gamma_{c,t}&=&\pm e^{i(\phi_\alpha-\phi_\beta)}
\left[n_A u_A^2 \cos^4{\theta\over 2}+n_B u_B^2 \sin^4{\theta\over 2}\right]
\label{gamma'}
\nonumber \\
&=&\pm e^{i(\phi_\alpha-\phi_\beta)}\times
n u_S^2 \left[\cos^4{\theta\over 2}+\sin^4{\theta\over 2}\right]
\end{eqnarray}
Notice that Eqs. (\ref{BSE_ct},\ref{BSE_1}) are always valid, whereas here
$\Pi\simeq \tau_S'(1-\tau_S' Dq^2)$.
Eqs. (\ref{tauS'}, \ref{gamma'}) suggest that in contrast to the topological mass case,
the scattering time (the self energy) and the bare vertex function (times $\Pi$) 
always cancel identically (giving unity) at the leading order of Eq. (\ref{BSE_1}).
As a consequence, the Cooperon diagram shows $1/q^2$ singularity,
which occurs always at the $l=1$-channel:
\begin{equation}
\Gamma_{t}={- e^{i(\phi_\alpha-\phi_\beta)} \over 2\tau_S'^2 Dq^2}.
\end{equation}
$\Gamma_{t}$ is indeed {\it positive and singular, independent of $\theta$},
indicating weak localization whenever the Fermi level is above the gap.

We have thus seen a clear distinction between two types of mass terms
in their weak localization properties, as $K$-$K'$ scattering is switched on.
In the ordinary mass case, the system shows 
the orthogonal
behavior, in sharp contrast to the unitary behavior of topological mass case.
Such different weak localization properties due to different types of mass terms
can be understood as follows.
Recall that the topological mass term induces a finite 
$\sigma_{xy}$, if one picks up only one of the two real spin components.
That was a clear signature of broken TRS, leading to unitary behavior.
In the ordinary mass, on the other hand,
contributions to $\sigma_{xy}$ from each valley cancel,
i.e., $\sigma^{K}_{xy}+\sigma^{K'}_{xy}=0$.
Therefore, there remains no trace of broken TRS any more,
once two valleys are coupled by $K$-$K'$ scattering.
The ordinary mass term does preserve TRS, and as a result
one finds always a diffusion-type $1/q^2$ singularity, leading to a  
weak localization behavior.

\section{Rashba spin-orbit interaction}
\label{Rashba}
Rashba spin-orbit interaction $H_R$ is also an important factor
for characterizing the physical properties of 
the doped Kane-Mele model.
A finite Rashba term appears only when the system loses
its inversion symmetry along the $z$-axis (perpendicular to the 2D ``graphene layer").
Physically, such breaking of inversion symmetry can be introduced by
an asymmetric potential, e.g., when a graphene sheet is placed on a substrate.
If Rashba spin-orbit interaction is stronger than a critical value ($\lambda_R>2\Delta$),
the system actually has no topologically non-trivial phase
\cite{KM2}.
Here we suppose that Rashba spin-orbit interaction is not too strong, and the undoped system is 
still in the topologically non-trivial phase.
However, we show, in this section,
that in regard to weak localization properties of our doped system,
Rashba spin-orbit interaction is still a relevant perturbation, and changes 
the symmetry class of our system, as soon as it is turned on
(in real samples of finite size of order $L^2$, there will be a crossover at 
the strength of Rashba spin-orbit interaction of order $\lambda_R\sim 1/L$). 
This is because the Rashba spin-orbit interaction mixes real spin up and down.
Before switching on $\lambda_R\neq 0$, we did have real spin up and down,
but they were just there, and inactive.

In the presence of $H_R$, we can still work on a $4\times 4$- 
(instead of $8\times 8$-) matrix space associated with
$\vec{\sigma}$ 
and $\vec{s}$ 
since $\tau_z$ is diagonal in our basis.
Focusing on one of the two valleys, say, K,
one notices that the Rashba spin-orbit interaction couples
A$\downarrow$ and B$\uparrow$ only
(here, the up and down arrows refer to the real spin).
The total Hamiltonian in the $K$-valley reads,
\begin{equation}
H_K=\left(
\begin{array}{cccc}
-\Delta & p_x-ip_y &0 & 0\\
p_x+i p_y & \Delta & i\lambda_R & 0\\
0 & -i \lambda_R & \Delta & p_x-i p_y\\
0 & 0& p_x+i p_y & -\Delta
\end{array}
\right)
\end{equation}
where the inner $2\times 2$ structure refers to AB-spin,
whereas the outer $2\times 2$ block structure
is asoociated with the real spin.
Here, for the sake of simplicity, we consider only the case of Kane-Mele type
topological mass term:
$\Delta\sigma_z\tau_z s_z$.
After diagonalization, 
the resulting four energy bands 
are classified into two conduction, 
and two valence bands, which we will call, respectively, 
$u\pm$ and $d\pm$.
The two valence bands $E_{d\pm}$ are 
degenerate on their top:
the top position is always at $E=-\Delta$, unaffected by the Rashba term,
whereas the conduction bands $E_{u\pm}$ are split by $2\lambda_R$:
\begin{eqnarray}
E_{u\pm}&=&|\vec{p}_\pm|\pm\lambda_R/2
\nonumber \\
&=&\sqrt{p_x^2+p_y^2+(\Delta\pm\lambda_R/2)^2}\pm\lambda_R/2,
\nonumber \\
E_{d\pm}&=&-|\vec{p}_\pm|\pm\lambda_R/2
\nonumber \\
&=&-\sqrt{p_x^2+p_y^2+(\Delta\pm\lambda_R/2)^2}\pm\lambda_R/2.
\label{enR}
\end{eqnarray}
The Corresponding eigenvectors $|u\pm\rangle$ and $|d\pm\rangle$
can be conveniently parametrized by introducing a fictitious 3D momentum:
\begin{equation}
\vec{p}_\pm=\left(
\begin{array}{c}
p_x\\
p_y\\
\Delta\pm\lambda_R/2
\end{array}
\right),
\label{p+-}
\end{equation}
and associated polar angles:
\begin{eqnarray}
\cos\theta_\pm&=&{\Delta\pm\lambda_R/2 \over 
\sqrt{p_x^2+p_y^2+(\Delta\pm\lambda_R/2)^2}}, 
\nonumber \\
\cos\phi&=&{p_x \over \sqrt{p_x^2+p_y^2}}.
\label{theta+-}
\end{eqnarray}
Note that $\phi$ is actually common to all cases.
Thanks to the parameters introduced in Eq. (\ref{theta+-}),
the eigenvectors corresponding to the eigenvalue $E_{u,d\pm}$,
given in Eqs. (\ref{enR}) allow for the following compact representation:
\begin{eqnarray}
|Ku\pm \rangle &=& {1\over\sqrt{2}}\left(
\begin{array}{c}
e^{-i\phi}\sin{\theta_\pm\over 2}\\
 \cos{\theta_\pm\over 2}\\
\mp i \cos{\theta_\pm\over 2}\\
\mp i e^{i\phi} \sin{\theta_\pm\over 2}
\end{array}
\right),
\nonumber \\
|Kd\pm\rangle &=&{1\over\sqrt{2}}\left(
\begin{array}{c}
e^{-i\phi} \cos{\theta_\pm\over 2}\\
- \sin{\theta_\pm\over 2}\\
\pm i \sin{\theta_\pm\over 2}\\
\mp i e^{i\phi} \cos{\theta_\pm\over 2}
\end{array}
\right).
\label{udpm}
\end{eqnarray}
Note that 
a crucial difference here compared with the unitary limit ($\lambda_R=0$)
is that because of the Rashba coupling, we have a stronger constraint on the
choice of our basis, and as a results there is no Berry phase in the matrix elements
(see below).
In the following, we consider again a doped system (our Fermi level is 
in the conduction band), i.e., $E>\Delta-\lambda_R$.

\subsection{Long-range scatterers: weak localization}
% --- orthogonal behavior}
The long-range scattering impurities do not couple $K$ and $K'$ valleys,
and in a given valley do not distinguish A and B sites, i.e., 
its scattering potential is propotional to unity (no projection).
In the following we focus on the $K$-valley.

\subsubsection{$\Delta-\lambda_R<E<\Delta+\lambda_R$ case}
Let us first consider the case in which the Fermi level is close to the bottom
of the conduction band, $\Delta-\lambda_R<E<\Delta+\lambda_R$.
In this case only $|u - \rangle$-branch contributes to our diagrams.
With this remark, we will omit, in the following, the indices $u$ and $-$, 
for specifying a ket such as $|Ku-\alpha\rangle$, and denote it simply as
$|K\alpha\rangle$.
As we have seen in Eqs. (\ref{udpm}),
because of the Rashba coupling, we have now a stronger constraint on the
choice of our basis, and as a results there is no Berry phase in the matrix element:
\begin{equation}
\langle K\beta| 1\otimes 1 |K\alpha\rangle = 
\cos (\phi_\alpha-\phi_\beta) \sin^2{\theta_-\over 2}+
\cos^2{\theta_-\over 2},
\label{noBerry}
\end{equation}
which is indeed {\it real}.
Note that
$|K\alpha\rangle$ here denotes $|Ku-\rangle$, given in Eqs. (\ref{udpm}),
for a fictitious 3D momentum $\vec{p_-}$, given in Eq. (\ref{p+-}),
and specified by $\alpha$.
Self-energy diagrams give the bare scattering time (without vertex correction) as
\begin{equation}
{1\over\tau_{L}} =\eta_L\left[
{1\over 2} \sin^4{\theta_-\over 2}+  \cos^4{\theta_-\over 2}.
\right]
\label{tauLR}
\end{equation}
Because of the $\cos (\phi_\alpha-\phi_\beta)$ term in Eq.(\ref{noBerry})
the bare (and also dressed) vertex function has several angular momentum
components.
Naturally, we expand them such that
\begin{eqnarray}
\gamma_{\alpha\beta}&=&\sum_l \gamma^{(l)}_{\alpha\beta} e^{i l(\phi_\alpha-\phi_\beta)},
\nonumber \\
\Gamma_{\alpha\beta}&=&\sum_l \Gamma^{(l)}_{\alpha\beta} e^{i l(\phi_\alpha-\phi_\beta)}.
\label{gamma_l}
\end{eqnarray}
The bare $\gamma_l$'s are given as,
\begin{eqnarray}
\gamma^{(0)}&=&\eta_L\left[ {1\over 2} \cos^4{\theta_-\over 2}+  \sin^4{\theta_-\over 2}\right],
\nonumber \\
\gamma^{(1)}&=&\gamma^{(-1)}={\eta_L\over 4}\cos^2 \theta_-, 
\nonumber \\
\gamma^{(2)}&=&\gamma^{(-2)}={\eta_L\over 4}\sin^4{\theta_-\over 2}.
\label{gamma--}
\end{eqnarray}
To solve the Bethe-Salpeter equations, first recall that 
there is an angular integration over the intermediate angle $\phi_\mu$,
which forbids coupling between $\Gamma^{(l)}$'s with different angular momentum $l$.
The Bethe-Salpeter equations take the form of Eq.(\ref{BSE_l}).
Then, comparing Eqs. (\ref{tauLR}) and (\ref{gamma--}),
one can verify that $\Gamma_0$ shows a diffusion-type
$1/q^2$-singularity, with a positive amplitude, leading to weak localization.

Recall that in the absence of Rashba spin-orbit interaction,
the anti-localization tendency toward the graphene limit was given by
the $l=1$ term. We emphasize that here the relevant (singular) contribution is 
from the $l=0$-channel, and this is quite contrasting to the former case.
We mentioned earlier that as Rashba spin-orbit interaction $\lambda_R\ne 0$,
with off-diagonal matrix elements in the real spin space, is turned on, 
the Berry phase, associated with the $1/q^2$ singularity of the
$l=1$-channel, disappears.
What we have discovered above is consistent with this observation.
It is also quite natural (as far as $K'$-valley is switched off) that the $1/q^2$ singularity appearing in the
$l=$(even) channel leads to  weak localization, 
whereas in the $l=$(odd) case, the same singularity leads to weak anti-localization.
From  a perspective point of view, it is also useful to remark  that
the number $N_s$ of effective spin degrees of freedom activated in the system
is increased from $N_s=1$ ($\lambda_R=0$) to $N_s=2$ $\lambda_R\ne 0$,
in switching on the Rashba term.
This aspect is summarized in TABLE II as explained later.
The parity of $N_s$ is a decisive factor for determining the symmetry class of
system.

\begin{figure}[!]
\includegraphics[width=7.5cm]{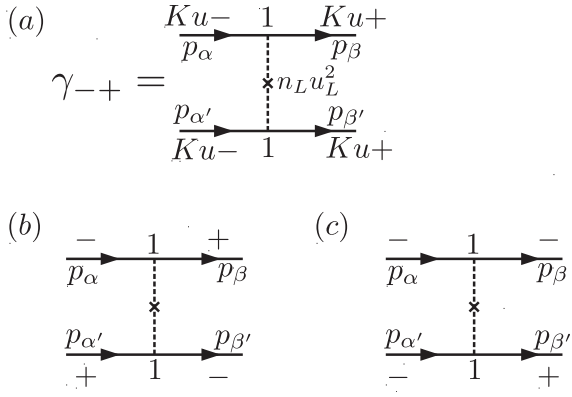}
\caption{Particle-particle ladders involving {\it inter-branch} processes.
In the presence of finite Rashba term $\lambda_R\ne 0$
(a) $\gamma_{-+}$ does contributes to $1/q^2$-singularity, 
whereas such diagrams as (b) and (c) are irrelevant to the singularity,
since $p_\alpha+p_{\alpha'}$ cannot be smaller than the order of
$\lambda_R$.}
\label{ladder-+}
\end{figure}

\subsubsection{$\Delta+\lambda_R<E$ case}
When the Fermi level is above the bottom of upper branch
of the conduction band, i.e., when $\Delta+\lambda_R<E$,
both $|Ku - \rangle$ and $|Ku + \rangle$ channels contribute to the weak localization 
properties. %which certainly introduces further complication.
In order to shorten the equations, we omit in this section even the index $K$ in
the brackets, and keep only $\pm$ for specifying branches,
e.g., $|-\alpha\rangle \equiv |Ku-\alpha\rangle$.
In this new notation,
\begin{eqnarray}
\langle -\beta| 1 |-\alpha\rangle &=&
\cos (\phi_\alpha-\phi_\beta) \sin^2{\theta_-\over 2}+
\cos^2{\theta_-\over 2},
\nonumber \\
\langle +\beta| 1 |-\alpha\rangle &=& 
-i\sin (\phi_\alpha-\phi_\beta) \sin {\theta_+\over 2} \sin {\theta_-\over 2}
\nonumber \\
&=&\langle -\beta| 1 |+\alpha\rangle,
\nonumber \\
\langle +\beta| 1 |+\alpha\rangle &=&
\cos (\phi_\alpha-\phi_\beta) \sin^2{\theta_+\over 2}+
\cos^2{\theta_+\over 2}.
\label{interb}
\end{eqnarray}
Using these matrix elements, one can calculate the scattering time $\tau_\pm$
for the $|Ku \pm \rangle$ branch: 
\begin{equation}
{1\over\tau_\pm}=\eta_L\left[{1\over 2}\sin^4{\theta_\pm\over 2}+\cos^4{\theta_\pm\over 2}+
{1\over 2}\sin^2{\theta_-\over 2}\sin^2{\theta_+\over 2}\right],
\end{equation}
where the last term corresponds to the contribution from inter-branch
matrix element, $|\langle +\beta| 1 |-\alpha\rangle|^2$.

The inter-branch matrix elements in Eqs. (\ref{interb}) also appear in the particle-particle ladders
(see FIG. 4).
In principle, the four electron states $\alpha$, $\beta$, $\alpha'$, $\beta'$ can belong to
either of the two channels, $|Ku - \rangle$ and $|Ku + \rangle$.
There is, however, an important simplification at this level.
Since we are interested only in the $1/q^2$ singular part of Cooperon diagrams,
we need $p_\alpha+p_{\alpha'}=q\simeq 0$, and similarly, 
$p_\beta+p_{\beta'}=q\simeq 0$.
This means that $\alpha$ and $\alpha'$ must belong to the same branch,
which is similarly the case for $\beta$ and $\beta'$.
We are thus led to consider such diagrams as, $\gamma_{--}$, $\gamma_{++}$,
$\gamma_{-+}$, and $\gamma_{+-}$.
$\gamma_{--}$ has already appeared in the regime, $\Delta-\lambda_R<E<\Delta+\lambda_R$
(its explicit form is also shown there, in the form of expansion with respect to $l$).
$\gamma_{-+}$ is defined in  FIG. 4 (a).
$\gamma_{+-}$ is similar to $\gamma_{-+}$, only with
$|Ku - \rangle$ and $|Ku + \rangle$ interchanged.
Other $\gamma$'s such as FIG. 4 (b) and (c) are irrelevant to $1/q^2$
singularity.
Explicit form of $\gamma$'s are given as,
\begin{eqnarray}
\gamma_{--}&=&  \eta_L \Big[ \cos^2 (\phi_\alpha-\phi_\beta) \sin^4 {\theta_-\over 2} 
\nonumber \\
&+& 2\cos^2 (\phi_\alpha-\phi_\beta) \sin^2 {\theta_-\over 2} \cos^2 {\theta_-\over 2}+
\cos^4 {\theta_-\over 2} \Big]
\nonumber \\
\gamma_{++}&=&  \eta_L \Big[ \cos^2 (\phi_\alpha-\phi_\beta) \sin^4 {\theta_+\over 2} 
\nonumber \\
&+& 2\cos^2 (\phi_\alpha-\phi_\beta) \sin^2 {\theta_+\over 2} \cos^2 {\theta_+\over 2}+
\cos^4 {\theta_+\over 2} \Big]
\nonumber \\
\gamma_{-+}&=& - \eta_L \sin^2 (\phi_\alpha-\phi_\beta) \sin {\theta_+\over 2} \sin {\theta_-\over 2}
\nonumber \\
\gamma_{+-}&=& \gamma_{-+}.
\end{eqnarray}
The dressed Cooperon diagrams, $\Gamma_{--}$ and $\Gamma_{+-}$, 
satisfy a coupled Bethe-Salpeter equation, which takes the following form:
\begin{equation}
\left(
\begin{array}{c}
\Gamma_{--}\\
\Gamma_{+-}
\end{array}
\right)%_{\alpha\beta}
=\left(
\begin{array}{c}
\gamma_{--}\\
\gamma_{+-}
\end{array}
\right)%_{\alpha\beta}
+\left(
\begin{array}{cc}
\gamma_{--}\Pi_-  &\gamma_{-+}\Pi_+ \\
\gamma_{+-}\Pi_- &\gamma_{++}\Pi_+
\end{array}
\right)%_{\alpha\mu}
\left(
\begin{array}{c}
\Gamma_{--}\\
\Gamma_{+-}
\end{array}
\right),
\label{BSE+-}
\end{equation}
where $\Pi_\pm \simeq \tau_\pm (1-D_\pm \tau_\pm q^2)$.
Recall that in the last term of Eq. (\ref{BSE+-}), combinations of the type,
$\gamma\Pi\Gamma$, appear, which implicitly contain averages over
the $\phi$-angle.
We, therefore, expand bare $\gamma$'s and dressed $\Gamma$'s, into different angular momentum
contributions, as Eqs. (\ref{gamma_l}),
and, to identify the singular contribution, pick up only the $l=0$ component. 
One way to convince oneself that
the dressed Cooperons show indeed $1/q^2$ singularity at the $l=0$ channel,
is to prove the following identity,
\begin{equation}
\det
\left(
\begin{array}{cc}
1-\gamma^{(0)}_{--}\tau_-  &- \gamma^{(0)}_{-+}\tau_+\\
-\gamma^{(0)}_{+-}\tau_-  &1-\gamma^{(0)}_{++}\tau_+
\end{array}
\right)=0.
\label{det0}
\end{equation}
In order to verify, first notice
\begin{eqnarray}
\gamma^{(0)}_{--}&=&\eta_L\left[ {1\over 2} \cos^4{\theta_-\over 2}+  \sin^4{\theta_-\over 2}\right],
\nonumber \\
\gamma^{(0)}_{-+}&=& - {\eta_L \over 2}\sin^2{\theta_-\over 2}\sin^2{\theta_+\over 2}
\nonumber \\
&=&\gamma^{(0)}_{+-}
\nonumber \\
\gamma^{(0)}_{++}&=&\eta_L\left[ {1\over 2} \cos^4{\theta_+\over 2}+  \sin^4{\theta_+\over 2}\right],
\end{eqnarray}
and use such relations as,
\begin{eqnarray}
1-\gamma^{(0)}_{--}\tau_-&=&(1/\tau_- -\gamma^{(0)}_{--})\tau_-
\nonumber \\
&=&\tau_- {\eta_L \over 2} \sin^2 {\theta_+\over 2} \sin^2 {\theta_-\over 2}
=-\tau_- \gamma^{(0)}_{+-},
\nonumber \\
1-\gamma^{(0)}_{++}\tau_+ &=& - \tau_+ \gamma^{(0)}_{-+}.
\end{eqnarray}

Based on these observations, we claim that
Rashba spin-orbit interaction
recovers the $1/q^2$ singularity of Cooperons, driving the system 
back to weak localization (orthogonal symmetry class),
whenever the Fermi level is above the gap.
These features are illustrated in FIG. 1.

Let us finally consider what happens if one adiabatically switches {\it off} the Rashba term.
The simplification we have made for justifying Eq. (\ref{BSE+-})
is no longer valid.
In the limit of vanishing $\lambda_R$ we cannot simply neglect such
diagrams as FIG. 4 (b), (c) and similar diagrams.
In principle, they could contribute equally to the $1/q^2$ singularity
if the singularity ever appears.
Relations such as $p_\alpha+p_{\alpha'}=q\simeq 0$ can become satisfied
in these diagrams.
At the same time, inclusion of all such diagrams makes the size of
coupled Bethe-Salpeter equation much bigger.
One can no longer decouple $\Gamma_{--}$ and $\Gamma_{+-}$
as Eq. (\ref{BSE+-}).
The Cooperons acquire more channels to couple to, and as 
a result, the cancellation property between particle-particle ladders and 
the self-energy, such as Eq. (\ref{det0}), is lost.
This loss of cancellation property due to activation of
those channels which are depicted in FIG. 4 (b) and (c)
explains why the system becomes unitary in the $\lambda_R=0$ limit.

On the other hand, it is also possible to choose a basis if $\lambda_R=0$, as we did in Sec. III,
in such a way that two upper band branches are decoupled.
Recall that in the absence of Rashba term,
the real spin part of the Hamiltonian is diagonalized in the real
spin basis (by diagonalizaing $s_z$).

\subsection{Short-range scatterers: weak anti-localization}

In the presence of Rashba spin-orbit interaction, we found weak localization behavior 
for long-range scatterers, on contrary to the graphene limit (compare the first (i) and third (iii) row of Table 1, 
left column).
Note also that $K$-$K'$ scattering drives the system, in the graphene limit, 
from symplectic to orthogonal symmetry class.
We thus finally consider short-range scatterers in the presence of Rashba interaction.
Since short-range scatterers involve inter-valley scattering, we need to consider also the eigenstates
at the $K'$-valley.
Diagonalizing the Hamiltonian at the $K'$-valley, one veryfies that
the energy spectrum is identical to Eq.(\ref{enR}); 
there are
two valence bands degenerate on their top at $E=-\Delta$, whereas
two conduction bands split by $2\lambda_R$.
Using the same parameterization as before, i.e.,
Eqs. (\ref{p+-}) and (\ref{theta+-}), 
the corresponding eigenvectors read,
\begin{eqnarray}
|K'u\pm \rangle &=& 
{1\over\sqrt{2}}
\left(
\begin{array}{c}
\cos{\theta_\pm \over 2}\\
- e^{-i\phi}\sin{\theta_\pm \over 2}\\
\mp i e^{i\phi} \sin{\theta_\pm \over 2}\\
\pm i \cos{\theta_\pm \over 2}
\end{array}
\right),
\nonumber \\
|K'd\pm \rangle&=&{1\over\sqrt{2}}\left(
\begin{array}{c}
 \sin{\theta_-\over 2}\\
e^{-i\phi} \cos{\theta_-\over 2}\\
\pm i e^{i\phi} \cos{\theta_-\over 2}\\
\pm i \sin{\theta_-\over 2}
\end{array}
\right).
\end{eqnarray}
Short range scatterers do distinguish AB pseudo-spin, whereas as far as they are non-magnetic impurities,
they are unconcerned about the real spin.
The impurity potential operator is proportional to ${\cal P}_{A,B}$,
which should be understood here as,
\begin{equation}
{\cal P}_{A,B}\otimes 1=\left(
\begin{array}{cc}
{\cal P}_{A,B}  &   0 \\
0  &   {\cal P}_{A,B} \\  
\end{array}
\right).
\end{equation}
Let us focus on the regime: $\Delta-\lambda_R< E<\Delta+\lambda_R$,
in which only $|Ku-\rangle$ and $|K' u-\rangle$ modes are available.
Using $\alpha,\beta,\cdots$ for specifying momenta, 
the relevant matrix elements read,
\begin{eqnarray}
\langle K'\beta | {\cal P}_A \tau_- |K\alpha\rangle&=&
{e^{-i\phi_\beta}+e^{-i\phi_\alpha}\over 4}
\sin{\theta_-},
\nonumber \\
\langle K \beta' | {\cal P}_A \tau_+ | K'\alpha' \rangle&=&
{e^{i\phi_{\beta'}}+e^{i\phi_{\alpha'}}\over 4}
\sin{\theta_-}
\nonumber \\
&\simeq&
-{e^{i\phi_\beta}+e^{i\phi_\alpha}\over 4}
\sin{\theta_-}.
\end{eqnarray}
As for the latter matrix element,
we noticed $\phi_{\alpha'}\simeq \phi_\alpha +\pi$,
$\phi_{\beta'}\simeq \phi_\beta +\pi$ in the second expression.
This clarifies the nature of additional minus sign
analogous to Eq. (\ref{add-}).
As for particle-particle ladders, there are, as before,
into $KK$, $KK'$-mixed, and $K'K'$ sectors.
We focus again on the $J=0$ sector, in which two types of bare vertex functions, 
$\gamma_\pm$,
are possible, and correspondingly, two types of Cooperons: 
$\Gamma_\pm$ (recall FIG. 2).
%Though only $\Gamma_{t}$ gives corrections to conductivity,
$\Gamma_\pm$ obey a coupled Bethe-Salpeter equation equation, (\ref{BSE+-}).
Again, as a result of projection ${\cal P}_A,{\cal P}_B$ and $K$-$K'$ scattering,
$\gamma_{t}$ acquires an additional minus sign:
\begin{equation}
\gamma_{t} = - {\eta_S\over 4}\cos^2 {\phi_\alpha-\phi_\beta \over 2}
\sin^2 \theta_-.
\label{gamma-R}
\end{equation}
However, in the present case, the Berry phase had
disappeared, implying that this time the new minus sign leads the system from orthogonal
to symplectic.
To verify this, first notice that 
intra-valley matrix elements are given as,
\begin{eqnarray}
\langle K \beta|{\cal P}_A |K \alpha\rangle&=&{1\over 2}\left[
e^{-i(\phi_\alpha-\phi_\beta)}
\sin^2{\theta_-\over 2}+
\cos^2{\theta_-\over 2}\right],
\nonumber \\
\langle K' \beta|{\cal P}_A |K' \alpha\rangle&=&{1\over 2}\left[
\cos^2{\theta_-\over 2}+
e^{i(\phi_\alpha-\phi_\beta)}
\sin^2{\theta_-\over 2}\right].
\end{eqnarray}
These combine to give,
\begin{eqnarray}
\gamma_{c} ={\eta_S\over 4} \left[
\cos^4{\theta_-\over 2}+
\cos (\phi_\alpha-\phi_\beta) {\sin^2 \theta_- \over 2}
+\sin^4{\theta_-\over 2}
\right],
\label{gamma+R}
\end{eqnarray}
which is indeed positive.
The scattering time is also calculated to be,
\begin{equation}
1/\tau_K=1/\tau_{K'}=\eta_S/4.
\label{tauSR}
\end{equation}
Note that all $\theta_-$-dependence cancelled, giving unity.

Let us now sum up the particle-particle ladders, and
then solve Bethe-Salpeter equation.
After diagonalization, one finds,
\begin{eqnarray}
\Gamma_{c}+\Gamma_{t}&\sim&({\rm regular}),
\nonumber \\
\left[1-(\gamma_{c}-\gamma_{t})\Pi\right](\Gamma_{c}-\Gamma_{t})&=&\gamma_{c}-\gamma_{t}>0.
\end{eqnarray}
In order to identify the singular contribution, one should 
expand $\gamma$'s and  $\Gamma$'s into different angular momentum components,
and pick up the $l=0$ term. Other contributions are indeed regular.
Comparing Eqs. (\ref{tauSR}), (\ref{gamma+R}) and (\ref{gamma-R}), one can indeed verify that
$\Gamma_{c}-\Gamma_{t}$ shows $1/q^2$-singularity, whereas
only $\Gamma_{t}$ contributes to the conductivity, giving  weak anti-localization correction.

Based on this observation, together with our analysis in the long-range case
for the regime: $E>\Delta+\lambda_R$,
one can naturally conjecture that this symplectic tendency
continues to the higher energy regime: $E>\Delta+\lambda_R$.
Then, the phase diagram in the presence of $K$-$K'$ scattering (FIG. 3),
becomes predominantly symplectic (when $\lambda_R\ne 0$).
Comparing two phase diagrams, in the absence (FIG. 1) and presence (FIG. 3) of $K$-$K'$ scattering,
one can see that the two weak anti-localization behaviors, one in the graphene limit, and the other of $Z_2$-topological
insulator phase, have a quite different origin.
The former occurs in a single Dirac cone, whereas, the latter occurs due to $K$-$K'$ scattering.
The former is related to Berry phase \`a la Ref.\cite{ANS}, whereas in the presence of Rashba
term mixing real spins, the matrix elements become real (the Berry phase disappears). 

Let us finally estimate the strength of gate electric field
required for observing the crossover to weak anti-localization.
For the crossover to be experimentally accessible,
Rashba SOI needs to be the order of $\sim 1$ K.
This corresponds to the electric field of order $\sim 1 $V/nm \cite{Min},
a value attainable in double-gated graphene devices \cite{Vanders}.
The crossover to weak anti-localization will be observed for a sample with insignificant ripples.  
A similar crossover due to Rashba SOI has been observed in another context
in InGaAs/InAlAs quantum well \cite{Nitta}.

\section{Summary and discussions}

%%%%%%%%%%%%%%%%%%%%%%%%%%%%%%%%%%%%%%%
\begin{table*}[htdp]
\caption{Summary: weak localization properties of the Kane-Mele model.
The weak localization (WL) or weak anti-localization (AL) 
results depending on the presence or absece of inter-valley scattering, 
and Rashba spin-orbit interaction. 
Notice the role of $N_s$, the number of {\it activated} (pseudo) spin species.
The value of angular momentum $l$ is also indicated for the singular ($\sim 1/q^{2}$) 
Cooperon channel (see Eq. (\ref{gamma_l})).
In the presence of $K$-$K'$ scattering, activation of the valley spin increases 
$N_s$ by one, leading to change of weak localization properties.} 
\begin{center}
\begin{tabular}{l|l|l}
\hline\hline
&
long-range scatterers 
&
short-range scatterers 
\\
&
(no $K$-$K'$ scattering)
&
($K$ and $K'$ points coupled)
\\ \hline \hline
(i) graphene
&
$\vec\sigma$: $N_s=1$ (odd)
&
$\vec\sigma$, $\vec\tau$: $N_s=2$ (even)
\\
%$H_1=$ (2D Dirac)
&
$1/q^{2}: l=1 \rightarrow$ AL
&
$1/q^{2}: l=1 \rightarrow$ WL
\\ \hline
(ii) finite mass system
&
$\vec\sigma$: $N_s=1$ (odd)
&
$\vec\sigma$, $\vec\tau$: $N_s=2$ (even)
\\
(a) ordinary 
&
unitary (WAL as $E\rightarrow\infty$)
&
(a) $1/q^{2}: l=1 \rightarrow$ WL
\\
(b) topological 
&
&
(b) unitary (WL as $E\rightarrow\infty$)
\\ \hline
(iii) doped $Z_2$ insulator
&
$\vec\sigma$, $\vec s$: $N_s=2$ (even)
&
$\vec\sigma$, $\vec\tau, \vec s$: $N_s=3$ (odd)
\\
&
$1/q^{2}: l=0 \rightarrow$ WL
&
$1/q^{2}: l=0 \rightarrow$ AL
\\
\hline\hline
\end{tabular}
\end{center}
\label{summary}
\end{table*}
%%%%%%%%%%%%%%%%%%%%%%%%%%%%%%%%%%%%%

We have studied localization properties of the doped Kane-Mele model.
We assumed that the disorder is weak enough to apply the standard weak localization theory, i.e.,
we have focused on the leading order $1/g$ correction in the $1/g$ expansion of $\beta (g)$.
We have considered a phase diagram in the 
($E,\lambda_R$)-plane together with 
inverse lifetime of a Cooperon, $1/\tau$, in order to indicate the crossover behavior in the unitary case.
When $\tau\rightarrow \infty$, the corresponding Cooperon mode shows $1/q^2$-singularity
and gives  correction to the $\beta$-function. Otherwise there is no correction to the $\beta$-function
at the $1/g$ level and in the $L\rightarrow \infty$ limit.
In the presence of topological mass term,
the system shows predominantly a unitary behavior,
as expected from symmetry consideration.
However, as we have seen in Sec. III, 
the role of mass term to system's localization property
is a more subtle issue,
depending on the type of impurities, and the presence or absence
of Rashba spin-orbit interaction.

A broader perspective is obtained 
on the phase diagram of localization properties in terms of
the number $N_s$ of {\it activated} (pseudo) spin degrees of freedom.
TABLE \ref{summary} summarizes the weak localization (WL) properties of doped Kane-Mele model.
It shows that contrasting behaviors are closely related to the {\it parity} of $N_s$.
To be explicit, if $N_s$ is even, the system shows 
WL, whereas if $N_s$
is odd, the system undergoes weak anti-localization (AL).
The unitary behavior, on the other hand, is independent of the parity of $N_s$, and 
emerges whenever the effective
time-reversal symmetry, i.e., TRS or PTRS, is broken.
Thus, in order to identify the weak localization symmetry class, 
it is most important to count
active spin degrees of freedom, and whether
the effective 
time reversal symmetry is preserved.
Note that the pseudo-spin $\vec\sigma$ is odd under PTRS, but even under the genuin TRS.  Therefore the ordinary mass term in the third column does not break the TRS, and the case (a) shows the orthogonal behavior.

%%%%%%%%%%%%%%%%%%%%%%%%%%%%%%%%%%%%%
\begin{table}[htdp]
\caption{(Pseudo) time reversal operations $T_{\Sigma}$, 
relevant in the subspace spanned by activated spins.
Transformation property of a mass term 
${\cal O}=m\sigma_z,\ \Delta\sigma_z\tau_z s_z$
under $T_{\Sigma}$:
$
T_{\Sigma} {\cal O} T_{\Sigma}^{-1}
={\pm \cal O}$.
The sign appears in the table.
$U$ refers to unitary class.}
\begin{center}
\begin{tabular}{c|c|c|c|c}
\hline\hline
activated spins &$ \vec{\sigma}$ & $\vec{\sigma}, \vec{\tau}$ & $\vec{\sigma}, \vec{s}$ 
& $\vec{\sigma}, \vec{\tau}, \vec{s}$
\\ \hline
relevant TRS operation
&$T_{\sigma}$ & $T_{\sigma\tau}$ & $T_{\sigma s}$ & $T_{\sigma\tau s}$
\\ \hline \hline
$\sigma_z$ &$-\rightarrow U$&+&$-\rightarrow U$&+
\\ \hline 
$\Delta\sigma_z\tau_z s_z$ &$-\rightarrow U$&$-\rightarrow U$&+&+
\\ \hline \hline
\end{tabular}
\end{center}
\label{PTRS}
\end{table}
%%%%%%%%%%%%%%%%%%%%%%%%%%%%%%%%%%%%%

In order to distinguish active and inactive spins, 
we introduce (pseudo) TRS operations $T_\Sigma$, 
defined in the subspace $\Sigma$, such that
\begin{equation}
T_\Sigma(H_1+H_R) T^{-1}_\Sigma 
= H_1+H_R,
\end{equation}
where
$\Sigma=\{\vec{\sigma}\}$, $\{\vec{\sigma}, \vec{\tau}\}$, $\{\vec{\sigma}, \vec{s}\}$,
$\{\vec{\sigma}, \vec{\tau}, \vec{s}\}$.  
Their explicit representations are,
\begin{align}
&T_{\sigma}=-i\sigma_y C, \quad
T_{\sigma\tau}=\tau_x C, \nonumber\\
&T_{\sigma s}=(-i\sigma_y)(-is_y)C, \quad
T_{\sigma\tau s}=\tau_x (-is_y) C, 
\end{align}
where $C$ is complex conjugation.
$T_{\sigma\tau s}$ represents the genuine TRS operation.
Effective TRS of the system is, therefore, determined by the transformation property of
the mass term (see TABLE \ref{PTRS}).
When a mass term is odd against TRS, the system shows the unitary behavior.
If some (pseudo or genuine) TRS exists in the system,
its weak localization property is determined by the number $N_s$ of the activated spin degrees of freedom.
One can verify 
$T_\Sigma^2=1$ if $N_s$ is even, whereas
$T_\Sigma^2=-1$ if $N_s$ is odd.
The former (latter) corresponds to the orthogonal (symplectic) class
in the random matrix theory \cite{D},
and leads to constructive (destructive) interference
between two scattering processes transformed from one to the other by $T_\Sigma$.

The above symmetry arguments allow for a slight generalization of our localization
phase diagram (see FIG. \ref{venn}).
Recall that in Sec. IV A, we have chosen the mass to be $H_2= - \Delta\sigma_z s_z$.
Results of this analysis are consistent with the symmetry consideration.
In FIG. \ref{venn}, we further conjecture based on the symmetry consideration that
a similar analysis for an ionic mass term $H_2= M\sigma_z$ leads to 
the absence of WL (unitary behavior).
Thus, the four minus signs in TABLE \ref{PTRS} correspond to
four unitary phases in FIG. \ref{venn}.
We also point out that in terms of symmetry, ripples plays the same role as the ionic mass.

We mention that the results of our analyses summarized in TABLE \ref{summary}
and FIG. \ref{venn}
are also relevant to {\it intrinsic} single valley systems \cite{Andrei},
such as the one realized in HgTe/CdTe quantum well \cite{Laurens}.
In the latter system, a single pair (Kramers pair) of Dirac cones appear, 
in contrast to graphene 
(cf. graphene has two pairs of Dirac cones; they appear at $K$ and $K'$ points).
Note that two Dirac cones (Kramers pair, with real spin $\uparrow$ and $\downarrow$),
described in the model introduced in Ref. \cite{Andrei},
have the {\it same} sign for their mass term,
implying (superficially)
the ordinary (ionic) mass case in our language.
However, they have also different chiralities, and 
by a simple linear transformation (exchange of rows and columns),
they can be mapped, and indeed corresponds to the topological mass case,
studied in Sec. IV A.
Our diagnosis (see TABLE \ref{summary}, and also FIG. \ref{venn})
suggests that the single-valley HgTe/CdTe system
shows a crossover from unitary to orthogonal (WL) symmetry class,
on activating the real spin degrees of freedom
by a (Rashba-like) off-diagonal interaction between 
real spin $\uparrow$ and $\downarrow$ sectors.

What is then essential to the AL behavior characteristic to the doped Kane-Mele model?
As can be guessed by the analysis relying on the number $N_s$ of active spin species, the combination of Rashba interaction and the $K$-$K'$ scattering makes $N_s=3$.  In other words, the AL results even in the absence of the topological mass term, or even in the case of the ordinary mass term.  In this sense, the AL is not a peculiar property of the doped $Z_2$ insulator.
The hallmark of the doped $Z_2$ insulator becomes manifest
in the absence of the Rashba interaction.
Namely, the robustness of the unitary behavior against the range of the disorder potential 
is a fingerprint of topological mass term.
The unitary behavior is closely related to the quantized spin Hall effect in the undoped case.

Finally, we comment on the role of 
the channel index $l$ which 
is a quantum number associated with a relative momentum
of electrons before and after the scattering by an impurity.
See Eqs.(\ref{gamma012}) and (\ref{gamma_l}) for its definition.
As we have already seen in the body of the paper,
to identify the singular channel $l$ greatly helps to  
determine the weak localization property.
In the case of long-range scatterers, the two valleys are decoupled,
and we can safely focus on, say, the $K$-valley.
Then, in Sec. III and IV, we have seen that 
the $1/q^2$ singularity appearing in the $l=$(even) channel 
leads to WL,
whereas in the $l=$(odd) case, the same singularity leads 
to AL,
This situation is reversed in the presence of $K$-$K'$ scattering:
even $l$ leads to AL and odd $l$ to WL according to TABLE II.

The above selection rule on the singular channel $l$ is superficially
dependent on the choice of the spin part of the wave function.
In Sec. II and III, we have chosen to use a single-valued two-component spinor 
for describing AB sub-lattice spin eigenstates,
which were naturally extended to the single-valued four-component spinor
in the presence of Rashba spin-orbit interaction.
If one choose a double-valued spinor instead of Eq.(\ref{pseudo}),
pretending that the AB-sublattice spin realizes a real spin,
then the selection rule is shifted by one.
This shift, however, is the same for all cases in TABLE II.  Hence
if we take the difference of $l$ relative to the graphene limit, for example, the WL or
AL behavior does not depend on the choice of the spin eigenstates.

As we have seen in Secs. III and IV, the value $l$ of
singular channel does not change in the presence of $K$-$K'$ scattering, 
The value of $N_s$ is, on the other hand, increased by one,
in switching on the $KK'$ valley spin $\vec\tau$.
Since $\vec\tau$ flip does not involve $\phi_\alpha$ variables,
the value $l$ of the singular channel remains the same.
However, the nature of the channel does change by the $K$-$K'$ scattering.
Namely, switching on of the valley spin appears as an additional minus sign 
in particle-particle ladders.

\vspace{0.5cm}
\indent
In conclusion, we have constructed a poor man's phase diagram of the
doped and disordered Kane-Mele model.
In order to characterize the $Z_2$ nature of the model,
we considered a doped case in contrast to 
the more familiar topological insulator phase.
We also 
characterized the $Z_2$ nature from its {\it bulk} properties, 
instead of its {\it edge} properties, to which the $Z_2$ nature
is often attributed. \cite{KM1,SCZ}
Our analysis is restricted to the weak (anti-)localization level, 
and can be used for constructing an effective $\sigma$-model
description of the same model.
The basic mechanism discussed here will be also useful for the understanding
of different types of topological insulators. 

\acknowledgments
KI and KN acknowledge support from KAKENHI:
Grant-in-Aid for Young Scientists (B) 
KI: 19740189, KN: 20740167.


\begin{thebibliography}{99}
\bibitem{Delta}
The sign of topological mass is taken to be opposite to the original 
Kane-Mele opaper \cite{KM1}.
This is only for later purpose of making phase diagrams more concise.
Physically, the sign of $\Delta$ is determined by the the sign of next-nearest neighbor imaginary 
hopping, which is irrelvant to the localization properties.
\bibitem{HLN+}
A similar circumstance occurs in the convetional analysis of weak anti-localization \`a la
Ref. \cite{HLN}. When one considers scattering due to spin-orbit coupling, 
$\vec{l} \cdot \vec{s}$ type isotropic term leads, as is well-known, to weak anti-localization.
On the contrary, if one keeps only $l_z s_z$ component, then the system is known to stay,
in this case, orthogonal (in our case the system stays unitary when $\lambda_R=0$).
\end{thebibliography}
\end{document}